\documentclass[sigconf]{acmart}

\def\BibTeX{{\rm B\kern-.05em{\sc i\kern-.025em b}\kern-.08emT\kern-.1667em\lower.7ex\hbox{E}\kern-.125emX}}
    
\acmConference[FATES on the Web 2019]{}{May 13--14, 2019}{San Francisco, CA}

\begin{document}

\newcommand{\bill}[1]{{\texttt{\color{green} Bill: [{#1}]}}}
\newcommand{\luke}[1]{{\texttt{\color{blue} Luke: [{#1}]}}}

%
\title{In Defense of Synthetic Data}

%
\author{Luke Rodriguez}
\email{rodriglr@uw.edu}
\affiliation{%
  \institution{University of Washington Information School}
  \city{Seattle}
  \state{WA}
}

\author{Bill Howe}
\email{billhowe@uw.edu}
\affiliation{%
  \institution{University of Washington Information School}
  \city{Seattle}
  \state{WA}
}

%

%
\begin{abstract}

Synthetic datasets have long been thought of as second-rate, to be used only when ``real'' data collected directly from the real world is unavailable. But this perspective assumes that raw data is clean, unbiased, and trustworthy, which it rarely is.  Moreover, the benefits of synthetic data for privacy and for bias correction are becoming increasingly important in any domain that works with people.  Curated synthetic datasets --- synthetic data derived from minimal perturbations of real data --- enable early stage product development and collaboration, protect privacy, afford reproducibility, increase dataset diversity in research, and protect disadvantaged groups from problematic inferences on the original data that reflects systematic discrimination.  Rather than representing a departure from the true state of the world, in this paper we argue that properly generated synthetic data is a step towards responsible and equitable research and development of machine learning systems.

\end{abstract}

\maketitle

\section{Introduction}

As researchers in computer science continue to develop new methods and apply them to other domains, the nature and types of insights that we can make are shaped by the datasets to which we have access. Traditionally the trick has been to collect (or gain access to) some representative real-world dataset on the phenomenon that you wish to analyze or predict, and then use it to develop and train a state-of-the-art approach. With this approach has always come a baseline assumption that data sampled from the real world --- unaltered --- is the gold standard for algorithm development. In this paper, we challenge that assumption and explore common scenarios in which altering real datasets to produce a synthetic version can not only improve their usefulness, but prevent significant errors resulting from bias and protect the interests of individuals. Curated synthetic datasets derived from real data can be used for early-stage pipeline development and exploratory analysis, and can also ensure that systemic biases or unfairness in the raw data are not propagated to trained models.

By focusing on how researchers and policymakers interact with data in practice rather than focusing on properties of a fictional dataset a priori we can investigate other desirable properties of datasets.

\section{Benefits of synthetic data}

We explore the benefits of synthetic data directly, rather than making negative comparisons to the utility of other datasets. Each of the following subsections highlights a distinct benefit of this approach.

\subsection{Protecting privacy}
Synthetic data has long been used as a mechanism to protect privacy of the individuals represented in the data. In practice, aggregation and anonymization are assumed to be sufficient, but more recently research efforts have focused on achieving synthetic data with provable guarantees using differential privacy. While generating synthetic data of this type can be a delicate process \cite{dwork2009complexity}, algorithms to do so are readily available \cite{hardt2012simple}. Exploration of such datasets protects against leaking sensitive or harmful information about individuals.

\subsection{Enabling exploratory development}

As long as a dataset can offer insight to analysts or developers, there will be limitations to how widely it can be shared. Less sensitive (privacy-protected) synthetic data offers developers who often only need data for system design and testing an opportunity to start their work without access to the data their system will ingest once deployed. This data is more than sufficient for the purposes of tool design and debugging. Though this synthetic data loses some of the signal present in the original dataset, the encoded schematic information is what is relevant to this situation and the privacy risk incurred by using the unaltered data instead need not be taken on  \cite{howe2017synthetic}.

In practice, leveraging the privacy guarantees of synthetic data allows us to share data without being encumbered by data sharing agreements whose drafting and establishment might take months or even years \cite{young_beyond_2019}. Less sensitive data can be shared much more freely and without the same legal encumbrances that real datasets carry.

\subsection{Focusing CS research attention on problems of national priority}

Much of the research done in computer science relies on high-quality granular datasets for algorithm testing. These datasets also must be publicly available for publishing transparent and reproducible evaluations of algorithms, but this necessarily means that data from potentially sensitive domains cannot be used in this kind of research. This tension creates a fundamental disconnect between the kinds of problems analyzed in many machine learning publications (e.g. plant classification, income prediction, wine quality \cite{Dua:2017}) and problems of keen national and public interest (e.g. homelessness, mobility, education). 

Synthetic data allows us to produce datasets from these sensitive domains of interest, and helps computer science research not ``overfit'' to simplified, unrealistic application domains. Any loss of utility in the sense of individual-level signals that occurs in creating synthetic data does not diminish the utility of these datasets as algorithmic benchmarks. Algorithmic research focuses on evaluating the performance of the algorithms rather than reaching conclusions about the domain that the data is drawn from.

\subsection{Ensuring reproducibility}

In the current research environment, some research done in machine learning looks to address the problem of applicability by using datasets from sensitive domains. Such papers fail to be reproducible, however, as the data cannot be publicly shared. Synthetic data provides access to a more diverse set of data for research \cite{bellovin2018privacy}, while not compromising on the principle of reproducibility.
 
\subsection{Reducing multiple hypothesis issues}
Best practice in data exploration looks to mitigate false discoveries that come as a result of multiple hypothesis testing. In theory, this is done through specifying a test procedure before accessing the data. In practice, however, exploratory analysis of data is a much more fluid process, and questions are often asked in response to discoveries made in the process of analysis. To accommodate this, Dwork et al. have shown that a single holdout set can be safely reused to test multiple hypotheses as long as it is only accessed through differentially private queries \cite{dwork2015reusable}. Thus, a synthetic dataset generated according to differential privacy serves as a reusable holdout, as any queries made against the synthetic dataset are by guaranteed to be differentially private as a result of the data generation process.
 
\subsection{Avoiding propagation of bias}

Datasets will inevitably contain bias, whether for methodological reasons such as sampling or because of systemic discrimination or oppression. For example, historical data about incomes in the United States  will indicate that white males have more earning potential than females or minority groups. These are not desirable signals to have in datasets, and their inclusion risks the propagation of this discrimination when used to train algorithmic decision systems.

While methods have been developed recently to mitigate these biases in classification tasks \cite{kilbertus2017avoiding, nabi2018fair}, we argue that this is not sufficient. These approaches do avoid reifying biases through classification, but they do not protect against a data analyst or project manager discovering this signal in the data and internalizing it as meaningful. Thus we suggest that biased signals should be accounted for in the creation of a synthetic dataset \cite{feldman2015certifying, hardt2016equality}, which could then be used for exploratory analysis or classification.  

This approach to generating synthetic data does raise a significant problem: who gets to decide which relationships in a dataset are problematic and which ought to remain unchanged? We believe it is important to understand the role of synthetic data in a broader sociotechnical system, and the process of creating it should involve both domain experts intimately familiar with the problems at hand and stakeholders who represent the populations present in the data. But there are scenarios in which such a decision should not be controversial.  For example, when choosing synthetic data to support decisions that do not privilege white men above other groups. We intend that decisions about which biases to mitigate will be made in a transparent way, and that the conditions enforced on a dataset are made readily apparent to those who use it.
 
\subsection{Suppressing sensitive signals}

In practice, many datasets that we would hope to gain access to are collected by companies and other organizations who could have their own business practices or operating strategies reflected in the data they collect. In response, these companies invest heavily to fight attempts to release their data publicly.  But synthetic datasets offer a scalpel rather than a hatchet to protecting competitive advantage;  by treating the sensitive relationships as a form of bias in the original data, we can use the same techniques as used to avoid discrimination to create synthetic datasets that protect companies' sensitive information while releasing the rest of the data.
Empirically, failing to protect against the leak of this information means that the data will be entirely unavailable for research, despite any good faith interest in supporting research.

\section{Conclusion}

Synthetic datasets generated in order to protect privacy or to correct problematic signals will by definition differ from data as it was originally collected, but this does not imply inferiority or decreased utility. As a research community, we have an opportunity to establish the usefulness of synthetic data. A repository to collaboratively collect and manage data, equipped with algorithms for generating and curating synthetic versions would help make our community more FATE-aware. As we continue to think more critically about the role that our algorithms are playing in sociotechnical systems, it is imperative that we treat our data in the same light.

\bibliographystyle{ACM-Reference-Format}
\bibliography{sample-base}

\end{document}